\documentclass{elsart}
\usepackage{epsfig}
\begin{document}
\begin{frontmatter}
  \title{Empirical formula applied to the lowest excitation energies of the natural parity \\odd multipole states in even-even nuclei}
  \author{Guanghao Jin,}
  \author{Jin-Hee Yoon,}
  \author{Dongwoo Cha\corauthref{cor}}
  \corauth[cor]{Corresponding author. Fax: +82 32 866 2452.}
  \ead{dcha@inha.ac.kr}
  \address{Department of Physics, Inha University, Incheon
402-751, South Korea}
\begin{abstract}
We applied our recently proposed empirical formula, a formula quite successful in describing essential trends of the lowest excitation energies of the natural parity even multipole states, to the lowest excitation energies of the natural parity odd multipole states in even-even nuclei throughout the entire periodic table. Even though the systematic behavior of the lowest excitation energies of odd multipole states is quite different from those of even multipole states, we have shown that the same empirical formula also holds reasonably well for the odd multipole states with the exception of a few certain instances.
\end{abstract}

\end{frontmatter}

Nuclear physics can be explained from two standpoints. One is the portrait description where the excited states of an individual nucleus are studied in depth. The traditional nuclear structure models, which have contributed significantly to the development of nuclear physics, belong to this category. The other is the landscape description where some particular nuclear properties are examined in terms of simple nuclear variables over the wide span of the chart of nuclides. A well-known example of a landscape description is the Weizs\"{a}cker's semi-empirical mass formula first proposed in the early 1930's \cite{Weizsacker}. It can reproduce the binding energy of the ground state of nuclei very accurately throughout the entire chart of nuclides in terms of the mass number $A$ and the atomic number $Z$ with just a few fitted parameters. However, for a very long time, no expression of a similar sort has been available for the properties of excited states in nuclei until very recently when we proposed a simple empirical formula for the lowest excitation energies $E_x$ of the natural parity even multipole states in even-even nuclei \cite{Ha,Kim}. This formula, which depends only on the mass number $A$, the valence proton number $N_p$, and the valence neutron number $N_n$, is written as
\begin{equation} \label{GE}
E_x = \alpha A^{-\gamma} + \beta_p \exp ( - \lambda_p N_p ) +
\beta_n \exp ( - \lambda_n N_n )
\end{equation}
where $\alpha$, $\gamma$, $\beta_p$, $\beta_n$, $\lambda_p$, and $\lambda_n$ are six model parameters to be fixed from the data for each multipole. We have shown that this empirical formula is capable of explaining main trends of the lowest excitation energies $E_x$ of the natural parity even multipole states up to $10^+$ in even-even nuclei throughout the entire periodic table \cite{Kim}. In this work, encouraged by the apparent success of Eq.\,(\ref{GE}) in describing  essential trends of the lowest excitation energies $E_x$ of the natural parity even multipole states in even-even nuclei, we plan to test whether the same empirical formula can be successfully applied to the lowest excitation energies $E_x$ of the natural parity odd multipole states.

We start with reviewing the gross behavior exposed by the data on the lowest excitation energies $E_x$ of the natural parity states in even-even nuclei. The lowest excitation energies $E_x$ of the even multipole states including $2^+$ (empty circles), $4^+$ (solid circles), $6^+$ (empty triangles), $8^+$ (solid triangles), and $10^+$ (empty squares) are shown in Fig.\,\ref{fig-1}(a); while those of the odd multipole states including $3^-$ (solid circles), $5^-$ (empty triangles), $7^-$ (solid triangles), and $9^-$ (empty squares) are shown in Fig.\,\ref{fig-1}(b). Due to a reason we give shortly, the lowest excitation energies $E_x$ of the dipole states are shown separately by solid squares in Fig.\,\ref{fig-1}(c) on top of the same data shown in Fig.\,\ref{fig-1}(b). In these figures, the plotted points are connected by solid lines along the isotopic chains. The measured excitation energies $E_x$ for the lowest $2^+$ and $3^-$ states are cited from a compilation in Raman {\it et al}. \cite{Raman} and Kib\'{e}di {\it et al}. \cite{Kibedi}, respectively, and those for the other multipole states are extracted from the Table of Isotopes, 8th edition by Firestone {\it et al}. \cite{Firestone}. It is quite obvious, by comparing the two graphs shown in (a) and (b) of Fig.\,\ref{fig-1}, that there are similar aspects as well as a clear distinction in the overall shape of excitation energies $E_x$ between the even multipole and odd multipole states. The fact that the excitation energies $E_x$ are getting larger as the multipole of the state increases is still true for the odd multipole states as well as for the even multipole states. However, the shape of the odd multipole excitation energies $E_x$ as a whole is very much different from that of the even multipole excitation energies $E_x$ because the excitation energies of odd multipole states lie significantly closer together than those of the even multipole states. We can see also from Fig.\,\ref{fig-1}(c) that the lowest dipole excitation energies $E_x$ lie scattered over the region where the lowest excitation energies $E_x$ of other odd multipole states are located as shown in Fig.\,\ref{fig-1}(b) rather than where they are located below the lowest $3^-$ excitation energies. This is the reason why we show the dipole excitation energy graph in addition in a separate panel (c) of Fig.\,\ref{fig-1}.

Now, in Fig.\,\ref{fig-2}, we plot once again the excitation energies $E_x$ of exactly the same set of lowest natural parity states previously shown in Fig.\,\ref{fig-1}.  This time, however, the excitation energies $E_x$ are not taken from the data but calculated using the empirical formula, Eq.\,(\ref{GE}), with the parameter set given in Tab.\,\ref{tab-1}. The parameter values for the even multipole states presented in Tab.\,\ref{tab-1} were calculated in Ref.\,\cite{Kim}. Also, the graphs shown in Fig.\,\ref{fig-2}(a) are, in fact, the same as the graphs found in Figs. 1-5 of Ref.\,\cite{Kim}. But, the parameter values listed in Tab.\,\ref{tab-1} for the odd multipole states are determined in this work by exactly the same method explained in Ref.\,\cite{Kim}. We first take the logarithmic error $R_E(i)$, for the $i$th data point, of the calculated excitation energies $E_x^{\rm cal} (i)$ with respect to the measured excitation energies $E_x^{\rm exp} (i)$ by \cite{Sabbey}
\begin{equation} \label{error}
R_E(i) = \log \left[ { E_x^{\rm cal}(i) \over E_x^{\rm exp}(i)} \right] = \log \Big[ E_x^{\rm cal}(i) \Big] - \log \Big[ E_x^{\rm exp}(i) \Big].
\end{equation}
Then the parameter values are determined by minimizing the $\chi^2$ value which is defined in terms of the logarithmic error by
\begin{equation} \label{Chi}
\chi^2 = { 1 \over {N_0}} \sum_{i=1}^{N_0} \Big| R_E(i) \Big|^2
\end{equation}
where $N_0$ is the number of total data points considered. The graphs of the odd multipole states shown in Fig.\,\ref{fig-2}(b) and (c) are prepared by those parameter values. By comparing the even multipole graphs shown in Fig.\,\ref{fig-2}(a) with those shown in Fig.\,\ref{fig-1}(a), we find that the overall shapes of the former graphs are quite similar to those of the latter graphs. Furthermore, by comparing the odd multipole graphs shown in Figs.\,\ref{fig-2}(b) and (c) with those shown in Figs.\,\ref{fig-1}(b) and (c), we also observe that the shapes of the graphs of the odd multipole excitation energies calculated using Eq.\,(\ref{GE}) are quite similar to the shapes of the graphs that are determined by the data except the particular two regions which are indicated by the shaded areas in parts (b) and (c) of both Figs.\,\ref{fig-1} and \ref{fig-2}.

Let us look closely into the excitation energies $E_x$ of nuclei which belong to the shaded areas in Figs.\,\ref{fig-1} and \ref{fig-2}. For that purpose, we indicate in Fig.\,\ref{fig-3}, the measured excitation energies by filled triangles or circles while the calculated energies by empty triangles or circles. In addition, if the measured and calculated excitation energies are relatively close, they are marked by triangles and otherwise by circles. From Fig.\,\ref{fig-3}, we confirm clearly that the excitation energies denoted by circles deviate, by quite a significant amount, from the overall trend of the excitation energies imposed by the empirical formula for all of the lowest natural parity odd multipole states. Besides, we find the following very interesting point. The excitation energies $E_x$ marked by circles between $A=220 \sim 232$ belong to those nuclei whose proton number is equal to a number between 86 and 92; while the excitation energies $E_x$ marked by circles between $A=144 \sim 152$ belong to those nuclei whose neutron number is also equal to a number between 86 and 92. This allows us to conjecture that the nucleon numbers, equal to a number between 86 and 92, play a certain important role in the deviation from the overall trend observed in the calculated excitation energies $E_x$ of the lowest natural parity odd multipole states. Incidentally, we have been informed later that those nuclei, whose mass number lies between $A=144 \sim 152$ and between $A=220 \sim 232$, coincide with those in which the octupole deformation plays an important role \cite{Cottle,Nazarewicz}. The octupole deformation in nuclei is known to be produced by the long-range octupole-octupole interaction between nucleons \cite{Ahmad}. The octupole correlations depend on the $E3$ matrix elements between single particle states with $\Delta j = \Delta l = 3$ and the energy spacing between them. There exist particular shell structure arrangements which meet the above constraints and give rise to strong octupole correlations due to the proximity of the single particle energy levels. Three such examples include nuclei whose Fermi surface lies: (i) between $2g_{9/2}$ and $1j_{15/2}$ for $N \approx 134$, (ii) between $2f_{7/2}$ and $1i_{13/2}$ for $N$ or $Z \approx 88$, and (iii) between $2d_{5/2}$ and $1h_{11/2}$ for $N$ or $Z \approx 58$. It is very interesting to note that the nuclei between $A=220 \sim 232$ correspond to those whose neutron and proton orbits belong to the cases (i) and (ii), respectively, so that the octupole-octupole interaction becomes maximum. Also, the nuclei between $A=144 \sim 152$ correspond again to those whose neutron and proton orbits belong to the cases (ii) and (iii), respectively.

In order to check more closely how well the empirical formula works, we have plotted, in each of Figs.\,\ref{fig-4}-\ref{fig-8}, the excitation energies $E_x$ of the lowest natural parity odd multipole states including $1_1^-$, $3_1^-$, $5_1^-$, $7_1^-$, and $9_1^-$, respectively, in the even-even nuclei against the mass number $A$ ($A$-plot). The upper two panels of each of these figures show the measured excitation energies while the lower two panels of each of these figures show the results for the excitation energies calculated by our empirical formula. The plotted points of the left two panels of each of these figures are, in fact, exactly same as those shown in Fig.\,\ref{fig-1}(b) (upper panels) and in Fig.\,\ref{fig-2}(b) (lower panels). Only after the parameter values of Eq.\,(\ref{GE}) were obtained employing the experimental input data from Refs.\,\cite{Kibedi,Firestone}, we are informed of the Evaluated Nuclear Structure Data File (ENSDF) database which contains the most current evaluated nuclear structure information \cite{ENSDF}. In the right two panels of each of Figs.\,\ref{fig-4}-\ref{fig-8}, we show the additional excitation energies extracted from the ENSDF database by open circles on top of the same plotted points shown in the left two panels. Therefore, the comparison should be made between the left upper and lower panels and also between the right upper and lower panels of each of the figures. Observing the graphs in the left two panels, we find that the empirical formula can reproduce the overall shape of the measured excitation energies $E_x$ of the natural parity odd multipole states in even-even nuclei reasonably well if, as mentioned earlier, we exclude the two regions indicated by the small solid horizontal bars in Figs.\,\ref{fig-4}-\ref{fig-8}. Furthermore, it is very good to confirm, from the graphs in the right two panels, that the additional data points extracted from the ENSDF database also follow the general trend predicted by our empirical formula.

The performance of our empirical formula in reproducing the lowest excitation energies $E_x$ of the natural parity odd multipole states can be inspected by drawing a scatter plot of the calculated excitation energies $E_x^{\rm cal}$ as a function of the measured energies, $E_x^{\rm exp}$. In Fig.\,\ref{fig-9}, we show such a scatter plot of the lowest excitation energies for not only the odd multipole (left panels) but also the even multipole (right panels) natural parity states. In addition, the average $\bar{R}$ and the dispersion $\sigma$ of the logarithmic error $R_E$ defined by Eq.\,(\ref{error}) for both the odd and even multipole states are shown in Tab.\,\ref{tab-2}. As before, the scatter plot in Fig.\,\ref{fig-9}, the average $\bar{R}$, and the dispersion $\sigma$ in Tab.\,\ref{tab-2} for the even multipole cases are quoted from Ref.\,\cite{Kim}. By comparing the results for the odd multipole states with those for the even multipole states from Fig.\,\ref{fig-9} and Tab.\,\ref{tab-2}, we find that the performance of the empirical formula for odd multipole states is as good as, if not better, than that for the even multipole states. Moreover, we show, in the middle panels of Fig.\,\ref{fig-9}, a scatter plot of the lowest excitation energies included additionally from the ENSDF database. The recalculated $\chi^2$ values and the recounted total number $N_0$ of the data points after including the additional excitation energies from the ENSDF database are listed inside the parentheses of Tab.\,\ref{tab-1}. The new dispersion $\sigma$ of the total data points is also shown in the parentheses of Tab.\,\ref{tab-2}. All the above results for the additional excitation energies obtained from the scatter plot, the $\chi^2$ values, and the dispersion $\sigma$ make us to convince that the prediction power of the empirical formula is not so bad after all.

It is hard to expect that the empirical formula, like the one given by Eq.\,(\ref{GE}) where $E_x$ is determined purely phenomenologically employing only the mass number $A$ and the valence nucleon numbers, $N_p$ and $N_n$, can accurately reproduce the nuclear shell effects in nuclei in the vicinity of closed shells. In order to inspect this point, we redraw, in Fig.\,\ref{fig-10}, two of the scatter plots, particularly for the octupole states (left panel), among the left panels of Fig.\,\ref{fig-9}, and for the quadrupole states (right panel) among the right panels of Fig.\,\ref{fig-9}, where the calculated excitation energies $E_x^{\rm cal}$ are plotted as a function of the measured excitation energies $E_x^{\rm exp}$. In Fig.\,\ref{fig-10}, the plotted points are expressed by the following different symbols, according to what type of nuclei they belong to: solid circles (both of proton and neutron shells are closed); solid squares (only one of proton and neutron shells is closed); open circles (both of proton and neutron shells are not closed). Against our naive expectation, however, we only find from the scatter plot that the degree of the logarithmic dispersion from $E_x^{\rm cal} = E_x^{\rm exp}$ line is about the same regardless of shell structure of the nucleus to which a plotted point belongs. We attribute this peculiar results to the fact that the model parameters in Eq.\,(\ref{GE}) are determined simply by fitting all of the data points from nuclei throughout the entire periodic table on an equal footing.

Finally, we plot the same excitation energies $E_x$ of the natural parity odd multipole states as shown in Fig.\,\ref{fig-4}-\ref{fig-8} again in Fig.\,\ref{fig-11} but this time against the product $N_pN_n$ ($N_pN_n$-plot) instead of against the mass number $A$. For years, the $N_pN_n$-plot has attracted much interest because a very simple pattern emerges whenever the nuclear data concerning the lowest collective state such as the lowest quadrupole excitation energy, the quadrupole transition probability, and the quadrupole deformation parameter, and so on is studied \cite{Casten1}. Such a phenomenon has been called the $N_pN_n$ scheme in the literature \cite{Casten2}. In Fig.\,\ref{fig-11}(a) (left panels), we show the measured excitation energies while in Fig.\,\ref{fig-11}(b) (right panels), we show those calculated by the empirical formula. In these $N_pN_n$-plots, the excitation energies are depicted by the following different symbols, according to which proton major shell they belong to: solid triangles ($Z=2 \sim 28$); empty triangles ($Z=30 \sim 50$); empty and solid circles ($Z=52 \sim 82$); and empty and solid squares ($Z=84 \sim 100$). Particularly, the solid circles and solid squares in Fig.\,\ref{fig-11} represent the excitation energies of nuclei that do not follow the trend imposed by the empirical formula between $A=130 \sim 170$ and between $A=210 \sim 250$, respectively, as mentioned previously in the reference to Fig.\,\ref{fig-3}. By observing Fig.\,\ref{fig-11}(a), we find that the measured lowest excitation energies $E_x$ of the natural parity odd multipole states show almost the same simple pattern as the pattern obtained when the $N_pN_n$-plots of the lowest excitation energies $E_x$ of the natural parity even multipole states are drawn \cite{Kim}. Furthermore, by comparing the graphs in Fig.\,\ref{fig-11} (a) and (b), we also find that the empirical formula, Eq.\,(\ref{GE}), reproduces the measured pattern of the $N_pN_n$ plot quite closely except for the plotted points marked by solid circles and squares. And consequently, we can observe that the $N_pN_n$ scaling fails, in the case of the natural parity odd multipole states, for nuclei where the octupole deformation prevails.

In summary, we applied the same empirical formula, which was successful in describing the essential trend of the lowest excitation energies $E_x$ of the natural parity even multipole states in even-even nuclei, to the lowest excitation energies $E_x$ of the natural parity odd multipole states. We obtained the following three notable results. First, in spite of the fact that the gross behavior of the odd multipole lowest excitation energies is quite different from that of the even multipole states, we observed that the empirical formula could also describe the overall trend of the measured lowest excitation energies $E_x$ of the odd multipole states up to $9^-$. Furthermore, it turns out that the degree of the performance of the empirical formula for the odd multipole states is as good as that for the even multipole states according to the scatter plot (Fig.\,\ref{fig-9}) and and the dispersion $\sigma$ of the logarithmic error (Tab.\,\ref{tab-2}). Second, we also found that there exist two particular groups of nuclei for which the measured lowest excitation energies $E_x$ do not follow the overall trend imposed by the empirical formula as shown by Fig.\,\ref{fig-3}. Such a group of nuclei happens to be all the isotones whose neutron number is equal to a number between 86 and 92 or all the isotopes whose proton number is also equal to a number between 86 and 92. We emphasize that this kind of abnormality can be recognized only by a landscape description such as described in this work. Also, this abnormality may be regarded as a separate evidence of the existence of the octupole collectivity which has been known for a long time \cite{Ahmad}. Last, the $N_pN_n$-plots of Fig.\,\ref{fig-11} indicate that the lowest excitation energies $E_x$ of not only the natural parity even multipole states but also the natural parity odd multipole states complies with the $N_pN_n$ scheme. We find it quite interesting to note that, even though the $A$-plots of the lowest excitation energies $E_x$ for the natural parity odd multipole states are very distinct from those for the even multipole states, the $N_pN_n$-plots for the odd multipole states are very similar to those for the even multipole states.

\section*{Acknowledgements}
This work was supported by the Korea Science and Engineering Foundation(KOSEF) grant funded by the Korea government(MEST) (2006-8-0083).

\newpage
\section*{Figure Captions}
\noindent {\bf Figure 1}. The measured lowest excitation energies $E_x$ of the natural parity states in even-even nuclei \cite{Raman,Kibedi,Firestone}. The plotted points are connected by solid lines along the isotopic chains. Part (a) shows the lowest excitation energies $E_x$ of the even multipole states while part (b) shows those of odd multipole states except the dipole states. In part (c), the lowest excitation energies $E_x$ of the dipole states are displayed on top of the same data shown in part (b).

\begin{figure}[h]
\centering
\includegraphics[width=13.0cm,angle=0]{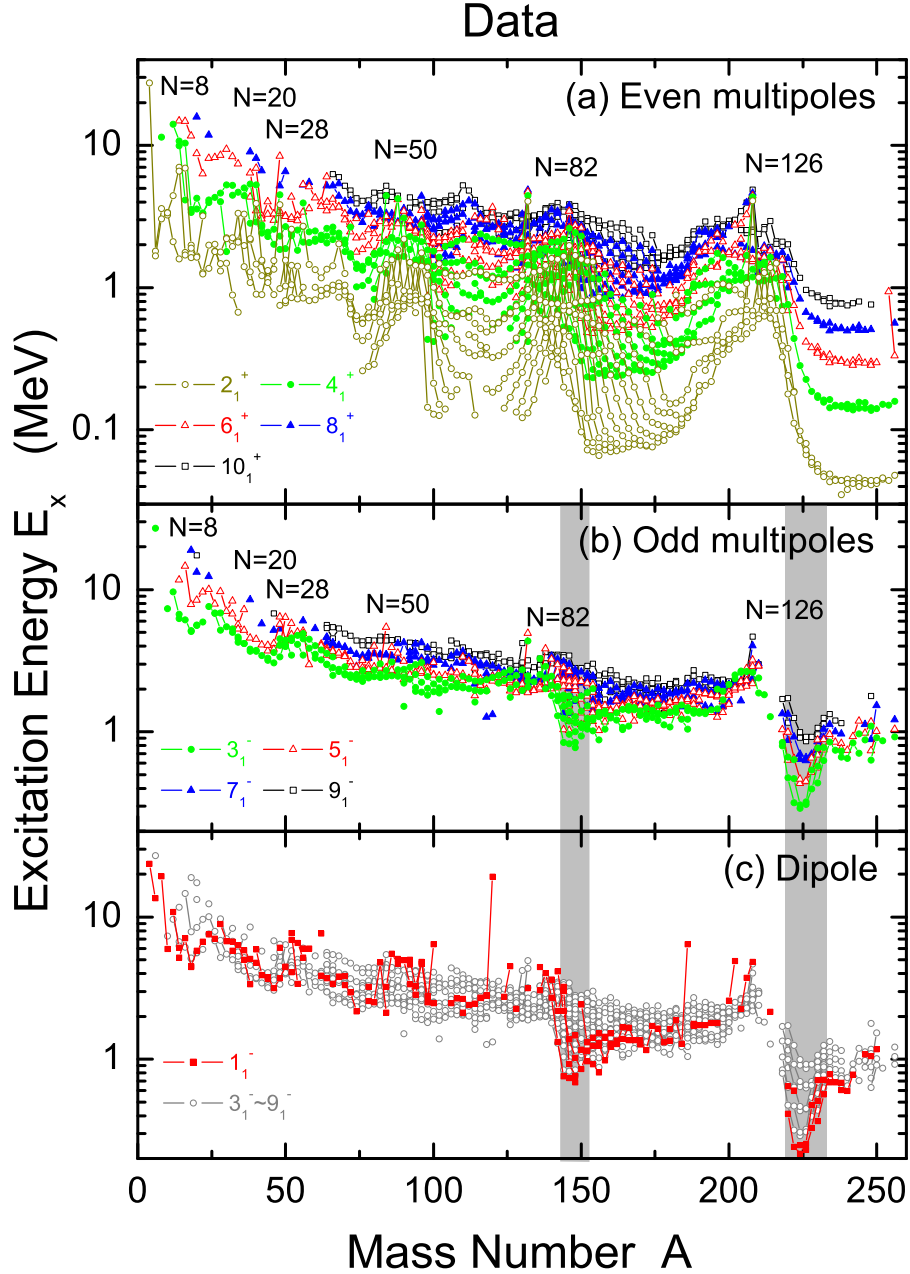}
\caption{The measured lowest excitation energies $E_x$ of the natural parity states in even-even nuclei \cite{Raman,Kibedi,Firestone}. The plotted points are connected by solid lines along the isotopic chains. Part (a) shows the lowest excitation energies $E_x$ of the even multipole states while part (b) shows those of odd multipole states except the dipole states. In part (c), the lowest excitation energies $E_x$ of the dipole states are displayed on top of the same data shown in part (b).}
\label{fig-1}
\end{figure}

\begin{figure}[h]
\centering
\includegraphics[width=13.0cm,angle=0]{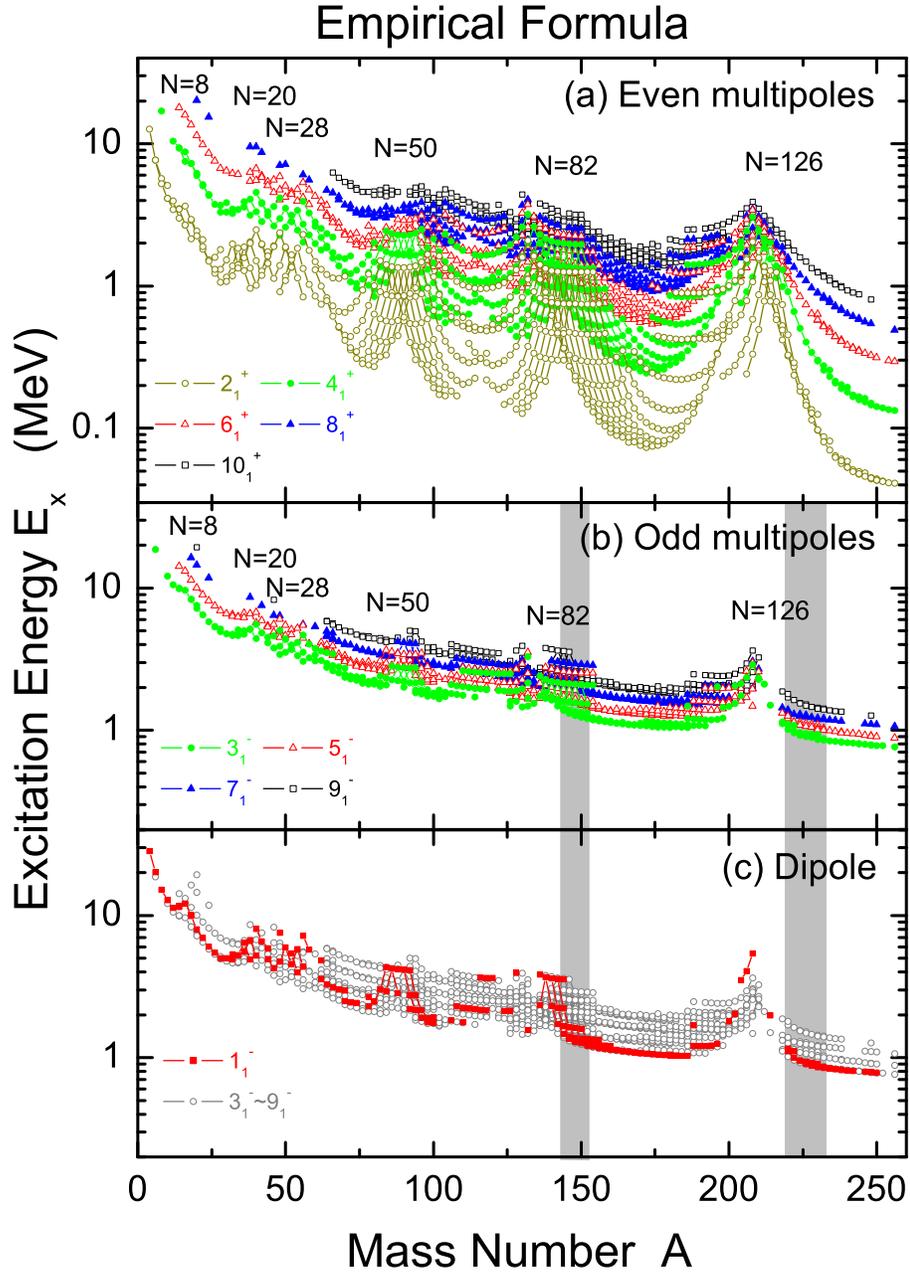}
\caption{The same as in Fig.\,\ref{fig-1} but for the excitation energies $E_x$ which are calculated by the empirical formula, Eq.\,(\ref{GE}), with the parameter set given in Tab.\,\ref{tab-1}.}
\label{fig-2}
\end{figure}

\begin{figure}[h]
\centering
\includegraphics[width=12.0cm,angle=0]{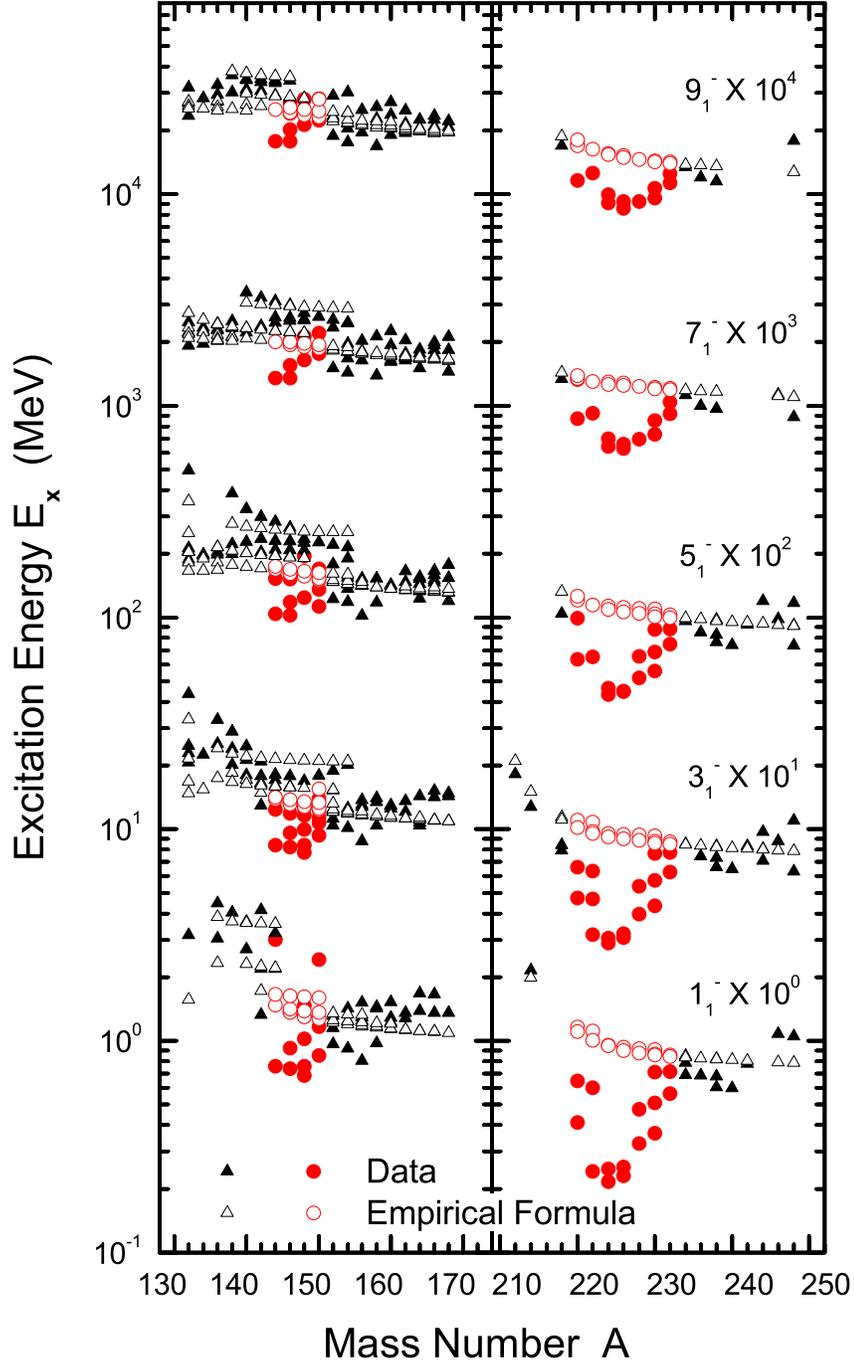}
\caption{The lowest excitation energies $E_x$ of the natural parity odd multipole states in even-even nuclei between $A=130 \sim 170$ and $A=210 \sim 250$. The filled triangles and circles denote the measured excitation energies while the empty ones show the excitation energies calculated by Eq.\,(\ref{GE}).}
\label{fig-3}
\end{figure}

\noindent {\bf Figure 2}. The same as in Fig.\,\ref{fig-1} but for the excitation energies $E_x$ which are calculated by the empirical formula, Eq.\,(\ref{GE}), with the parameter set given in Tab.\,\ref{tab-1}.

\noindent {\bf Figure 3}. The lowest excitation energies $E_x$ of the natural parity odd multipole states in even-even nuclei between $A=130 \sim 170$ and $A=210 \sim 250$. The filled triangles and circles denote the measured excitation energies while the empty ones show the excitation energies calculated by Eq.\,(\ref{GE}).

\noindent {\bf Figure 4}. The excitation energies $E_x$ of the lowest dipole states in even-even nuclei. The upper part shows the measured excitation energies while the lower part shows those calculated by Eq.\,(\ref{GE}). The measured excitation energies in the left upper panel are extracted from the Table of Isotopes, 8th edition by Firestone {\it et al}. \cite{Firestone}, while the additional excitation energies, which are extracted from the ENSDF database \cite{ENSDF}, are plotted in the right upper panel by open circles.

\begin{figure}[h]
\centering
\includegraphics[width=14.0cm,angle=0]{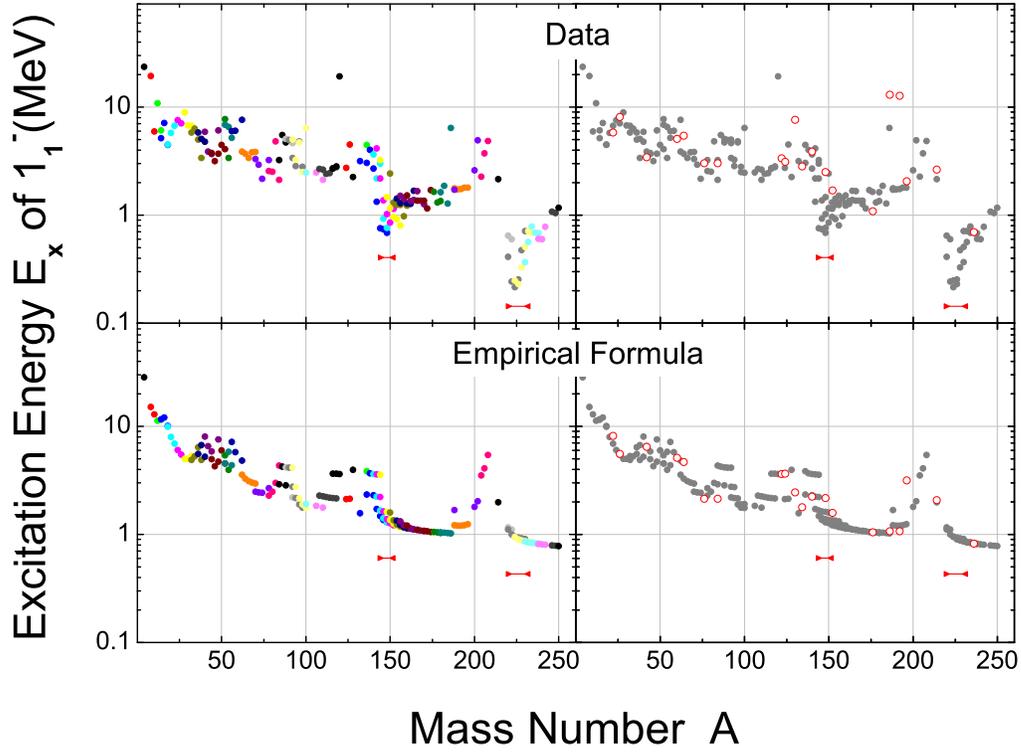}
\caption{The excitation energies $E_x$ of the lowest dipole states in even-even nuclei. The upper part shows the measured excitation energies while the lower part shows those calculated by Eq.\,(\ref{GE}). The measured excitation energies in the left upper panel are extracted from the Table of Isotopes, 8th edition by Firestone {\it et al}. \cite{Firestone}, while the additional excitation energies, which are extracted from the ENSDF database \cite{ENSDF}, are plotted in the right upper panel by open circles.}
\label{fig-4}
\end{figure}

\noindent {\bf Figure 5}. The same as in Fig.\,\ref{fig-4}, but for the excitation energies $E_x$ of the lowest $3_1^-$ states in even-even nuclei. The measured excitation energies in the left upper panel are quoted from the compilation in Kib{\' e}di {\it et al}. \cite{Kibedi}.

\noindent {\bf Figure 6}. The same as in Fig.\,\ref{fig-4}, but for the excitation energies $E_x$ of the lowest $5_1^-$ states in even-even nuclei. The measured excitation energies in the left upper panel are extracted from the Table of Isotopes, 8th edition by Firestone {\it et al}. \cite{Firestone}.

\noindent {\bf Figure 7}. The same as in Fig.\,\ref{fig-4}, but for the excitation energies $E_x$ of the lowest $7_1^-$ states in even-even nuclei. The measured excitation energies in the left upper panel are extracted from the Table of Isotopes, 8th edition by Firestone {\it et al}. \cite{Firestone}.

\noindent {\bf Figure 8}. The same as in Fig.\,\ref{fig-4}, but for the excitation energies $E_x$ of the lowest $9_1^-$ states in even-even nuclei. The measured excitation energies in the left upper panel are extracted from the Table of Isotopes, 8th edition by Firestone {\it et al}. \cite{Firestone}.

\noindent {\bf Figure 9}. The scatter plot of the calculated excitation energies $E_x^{\rm cal}$ as a function of the measured energies $E_x^{\rm exp}$ for the lowest excitation energies of the natural parity odd multipole (left and middle panels) as well as even multipole (right panels) states. The scatter plots shown in the middle panels employ only the additional excitation energies extracted from the ENSDF database \cite{ENSDF}. The graphs for the even multipole states are quoted from Ref.\,\cite{Kim}.

\noindent {\bf Figure 10}. Two of the scatter plots, particularly for the octupole states (left panel) among the left panels of Fig.\,\ref{fig-9}, and for the quadrupole states (right panel) among the right panels of Fig.\,\ref{fig-9} are redrawn. The plotted points are expressed by the following different symbols: solid circles (both of proton and neutron shells are closed); solid squares (only one of proton and neutron shells is closed); open circles (both of proton and neutron shells are not closed).

\begin{figure}[h]
\centering
\includegraphics[width=14.0cm,angle=0]{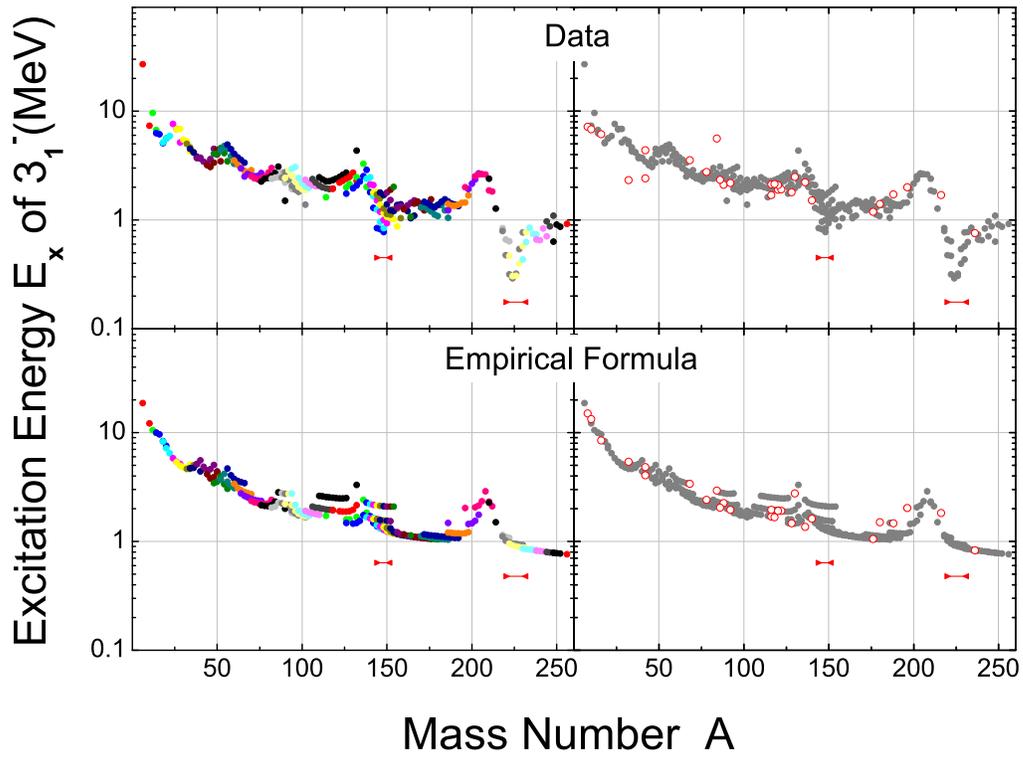}
\caption{The same as in Fig.\,\ref{fig-4}, but for the excitation energies $E_x$ of the lowest $3_1^-$ states in even-even nuclei. The measured excitation energies in the left upper panel are quoted from the compilation in Kib{\' e}di {\it et al}. \cite{Kibedi}.}
\label{fig-5}
\end{figure}

\begin{figure}[h]
\centering
\includegraphics[width=14.0cm,angle=0]{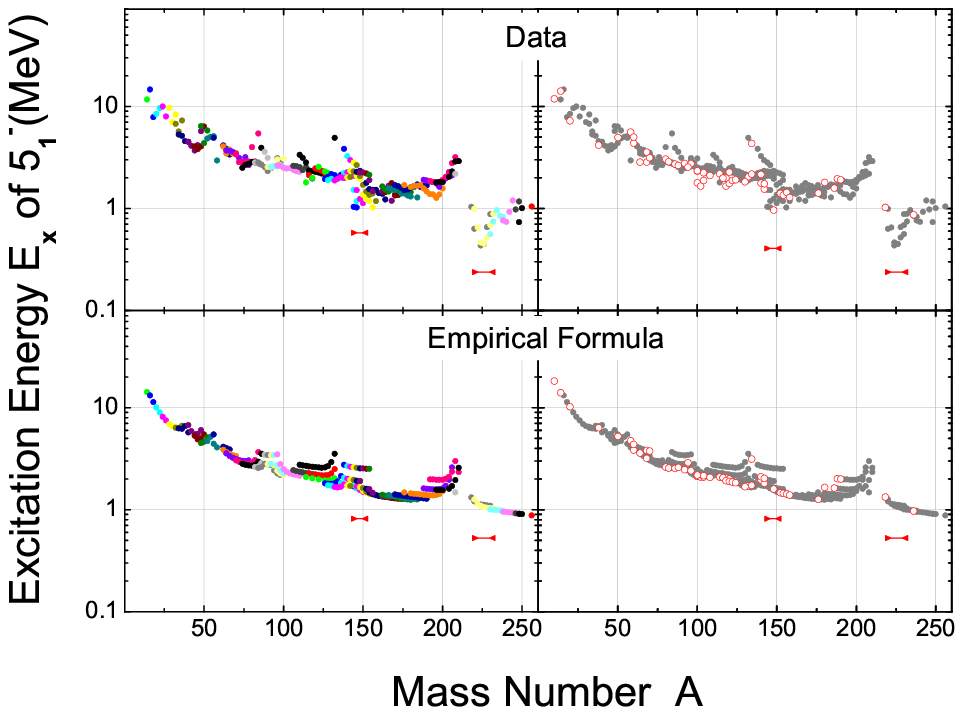}
\caption{The same as in Fig.\,\ref{fig-4}, but for the excitation energies $E_x$ of the lowest $5_1^-$ states in even-even nuclei. The measured excitation energies in the left upper panel are extracted from the Table of Isotopes, 8th edition by Firestone {\it et al}. \cite{Firestone}.}
\label{fig-6}
\end{figure}

\begin{figure}[h]
\centering
\includegraphics[width=14.0cm,angle=0]{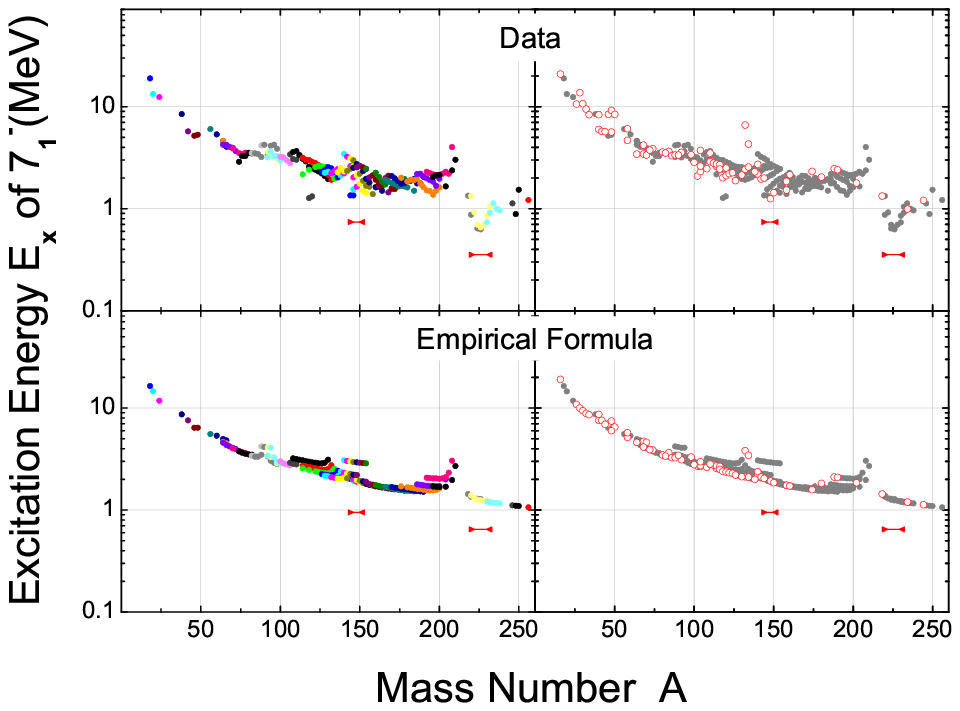}
\caption{The same as in Fig.\,\ref{fig-4}, but for the excitation energies $E_x$ of the lowest $7_1^-$ states in even-even nuclei. The measured excitation energies in the left upper panel are extracted from the Table of Isotopes, 8th edition by Firestone {\it et al}. \cite{Firestone}.}
\label{fig-7}
\end{figure}

\begin{figure}[h]
\centering
\includegraphics[width=14.0cm,angle=0]{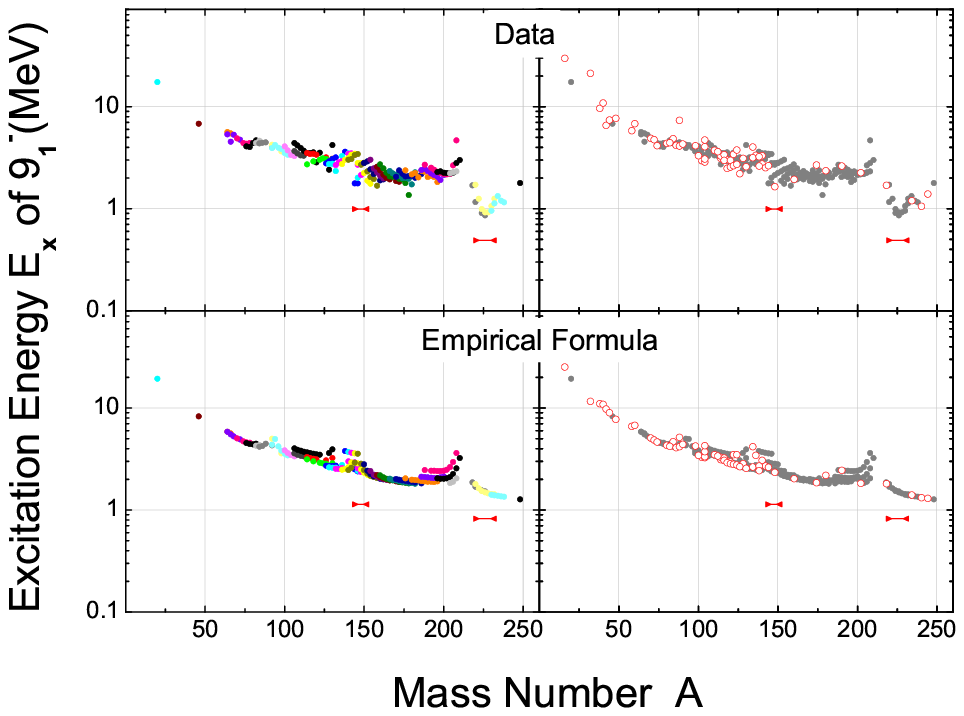}
\caption{The same as in Fig.\,\ref{fig-4}, but for the excitation energies $E_x$ of the lowest $9_1^-$ states in even-even nuclei. The measured excitation energies in the left upper panel are extracted from the Table of Isotopes, 8th edition by Firestone {\it et al}. \cite{Firestone}.}
\label{fig-8}
\end{figure}

\begin{figure}[h]
\centering
\includegraphics[width=11.0cm,angle=0]{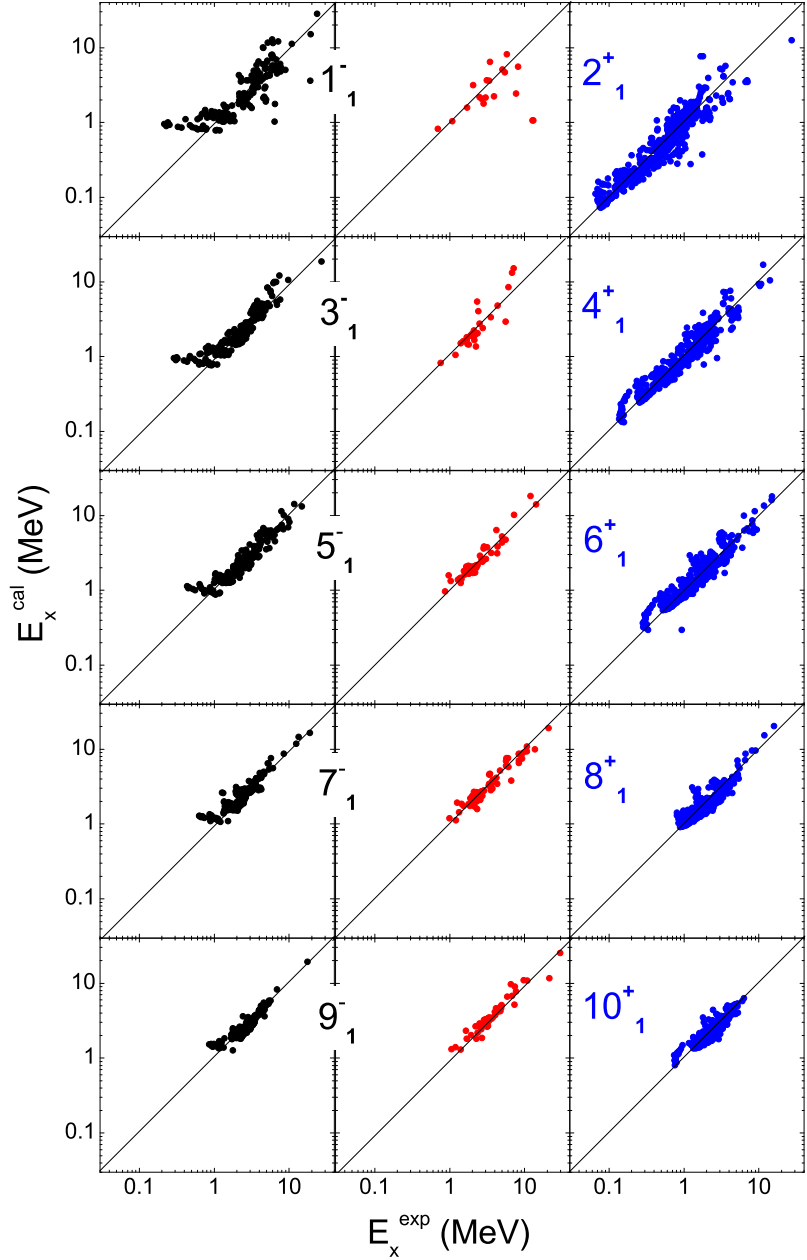}
\caption{The scatter plot of the calculated excitation energies $E_x^{\rm cal}$ as a function of the measured energies $E_x^{\rm exp}$ for the lowest excitation energies of the natural parity odd multipole (left and middle panels) as well as even multipole (right panels) states. The scatter plots shown in the middle panels employ only the additional excitation energies extracted from the ENSDF database \cite{ENSDF}. The graphs for the even multipole states are quoted from Ref.\,\cite{Kim}.}
\label{fig-9}
\end{figure}

\begin{figure}[h]
\centering
\includegraphics[width=14.0cm,angle=0]{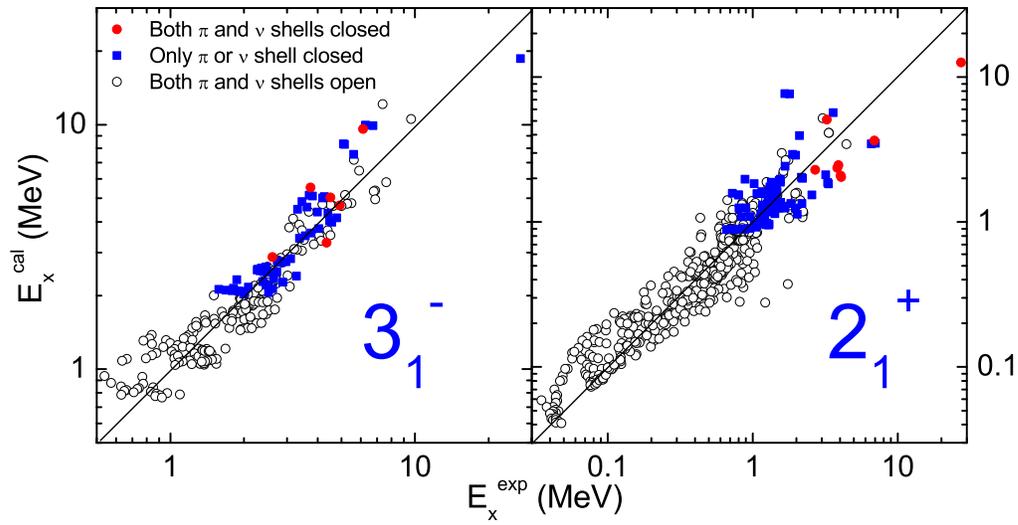}
\caption{Two of the scatter plots, particularly for the octupole states (left panel) among the left panels of Fig.\,\ref{fig-9}, and for the quadrupole states (right panel) among the right panels of Fig.\,\ref{fig-9} are redrawn. The plotted points are expressed by the following different symbols: solid circles (both of proton and neutron shells are closed); solid squares (only one of proton and neutron shells is closed); open circles (both of proton and neutron shells are not closed).}
\label{fig-10}
\end{figure}

\begin{figure}[h]
\centering
\includegraphics[width=14.0cm,angle=0]{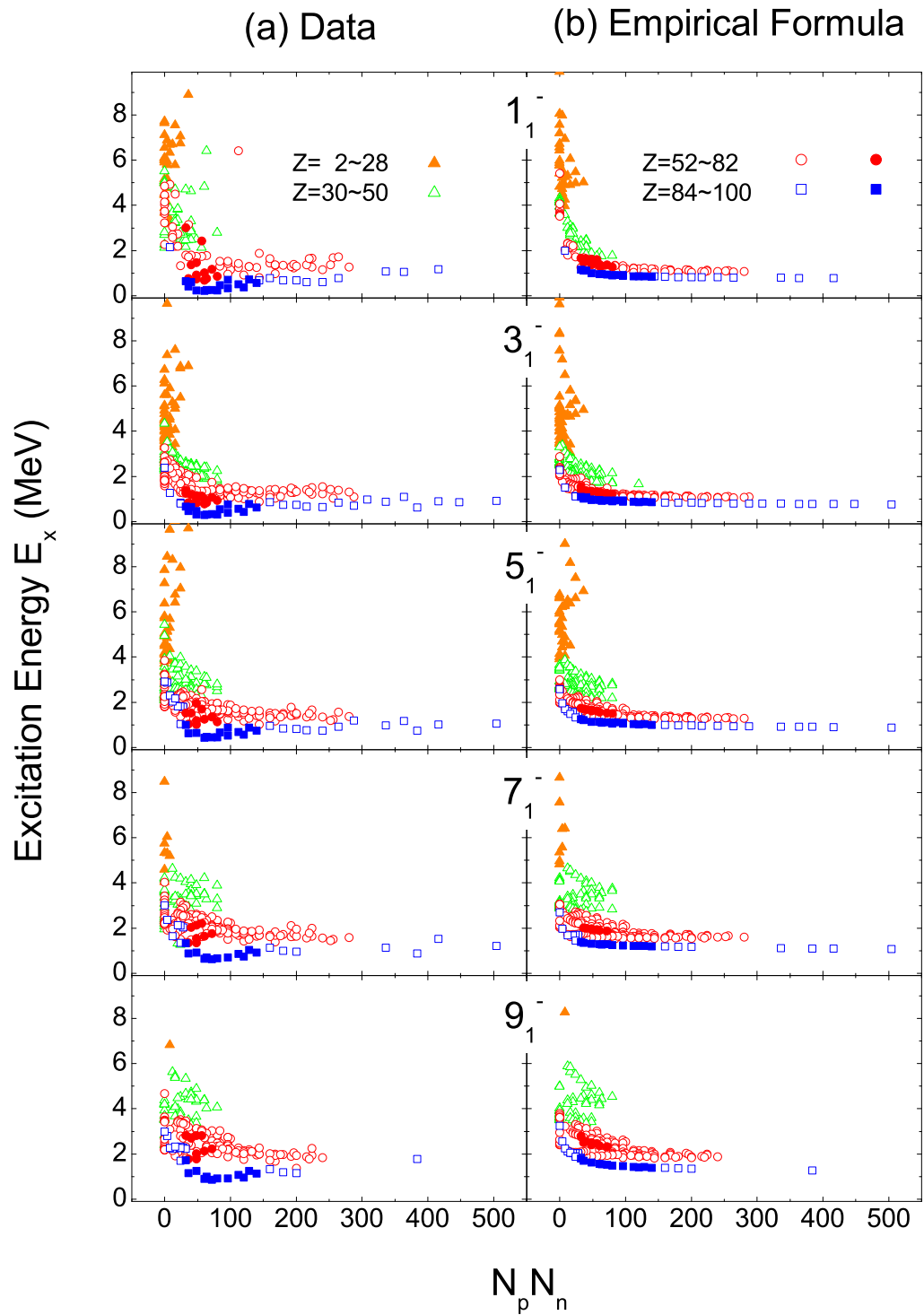}
\caption{The same as Figs.\,\ref{fig-4}-\ref{fig-8} but plotted against the product $N_pN_n$ instead of the mass number $A$.}
\label{fig-11}
\end{figure}

\noindent {\bf Figure 11}. The same as Figs.\,\ref{fig-4}-\ref{fig-8} but plotted against the product $N_pN_n$ instead of the mass number $A$.

\newpage
\begin{table}[h]
\begin{center}
\caption{The values adopted for the six parameters in Eq.\,(\ref{GE}) for the lowest excitation energies $E_x$ of the natural parity including even multipole $2_1^+$, $4_1^+$, $6_1^+$, $8_1^+$, and $10_1^+$ states and also odd multipole $1_1^-$, $3_1^-$, $5_1^-$, $7_1^-$, and $9_1^-$ states. The values for the even multipole states are quoted from Ref.\,\cite{Kim}. The last two columns are the $\chi^2$ value, defined by Eq.\,(\ref{Chi}), which fits the parameter set and the total number $N_0$ of the data points, respectively, for the corresponding multipole state. The numbers in the parentheses are obtained by including the additional excitation energies extracted from the ENSDF database \cite{ENSDF}.}
\begin{tabular}{ccccccccc}
\hline\hline
$J_1^\pi$~&~$\alpha$~&~$\gamma$~&~$\beta_p$~&
~$\beta_n$~&~$\lambda_p$~&~$\lambda_n$~&~$\chi^2$~
&~$N_0$~~\\
&(MeV)&&(MeV)&(MeV)&&&&\\
\hline
$2_1^+$&68.37&1.34&0.83&1.17&0.42&0.28&0.126&557\\
$4_1^+$&268.04&1.38&1.21&1.68&0.33&0.23&0.071&430\\
$6_1^+$&598.17&1.38&1.40&1.64&0.31&0.18&0.069&375\\
$8_1^+$&1,438.59&1.45&1.34&1.50&0.26&0.15&0.053&309\\
$10_1^+$&2,316.85&1.47&1.36&1.65&0.21&0.14&0.034&265\\
\hline
$1_1^-$&75.31&0.83&2.18&2.33&0.57&0.44&0.240(0.297)&177(196)\\
$3_1^-$&76.50&0.83&1.07&0.90&0.40&0.47&0.073(0.076)&289(317)\\
$5_1^-$&144.14&0.92&0.84&1.09&0.32&0.45&0.046(0.043)&297(352)\\
$7_1^-$&282.54&1.01&0.66&1.08&0.37&0.56&0.036(0.034)&241(315)\\
$9_1^-$&441.51&1.06&0.77&1.33&0.32&0.37&0.022(0.022)&204(267)\\
\hline \hline
\end{tabular}
\label{tab-1}
\end{center}
\end{table}

\newpage
\begin{table}[t]
\begin{center}
\caption{The average $\bar R$ and the dispersion $\sigma$ of the logarithmic error $R_E$ for the lowest excitation energies $E_x$ of the natural parity odd and even multipole states. The numbers in the parentheses are obtained by including the additional excitation energies extracted from the ENSDF database \cite{ENSDF}. The results for the even multipole states are quoted from Ref.\,\cite{Kim}.}
\begin{tabular}{cccccc}
\hline\hline
~~$J_1^\pi$~~&~~$1_1^-$~~&~~$3_1^-$~~&~~$5_1^-$~~&~~$7_1^-$~~&~~$9_1^-$~~\\
\hline
${\bar R} \times 10^6$&5&36&-6&6&2\\
$\sigma$&0.482&0.268&0.211&0.187&0.146\\
~&(0.535)&(0.273)&(0.205)&(0.182)&(0.147)\\
\hline \hline
$J_1^\pi$&$2_1^+$&$4_1^+$&$6_1^+$&$8_1^+$&$10_1^+$\\
\hline
${\bar R} \times 10^5$&-32&-83&-7&-591&152\\
$\sigma$&0.353&0.265&0.260&0.227&0.183\\
\hline \hline
\end{tabular}
\label{tab-2}
\end{center}
\end{table}

\newpage

\newpage

\newpage

\newpage

\newpage

\newpage

\newpage

\newpage

\newpage

\newpage

\newpage

\end{document}